\documentclass[twocolumn,showpacs,prl,aps,tightenlines,superscriptaddress,nofootinbib,eqsecnum,amsfonts,amsmath]{revtex4}
\usepackage{graphicx}
\def\be{\begin{equation}}
\def\en{\end{equation}}
\def\bea{\begin{eqnarray}}
\def\ena{\end{eqnarray}}

\begin{document}

\title{Resolving Super Massive Black Holes with LISA}

\author{Stanislav Babak}
\affiliation{Max-Planck-Institut f\"ur Gravitationsphysik, Albert-Einstein-Institut, Am M\"uhlenberg 1, D-14476 Golm bei Potsdam, Germany}
\author{Mark Hannam}
\affiliation{Physics Department, University College Cork, Cork, Ireland}
\author{Sascha Husa}
\affiliation{Max-Planck-Institut f\"ur Gravitationsphysik, Albert-Einstein-Institut, Am M\"uhlenberg 1, D-14476 Golm bei Potsdam, Germany}
\author{Bernard Schutz}
\affiliation{Max-Planck-Institut f\"ur Gravitationsphysik, Albert-Einstein-Institut, Am M\"uhlenberg 1, D-14476 Golm bei Potsdam, Germany}

\begin{abstract}
We study the angular resolution of the gravitational wave detector LISA
and show that numerical relativity can 
drastically improve the accuracy of position location 
for coalescing Super Massive Black Hole (SMBH) binaries. 
For systems with total redshifted mass above $10^7\,  M_{\odot}$, LISA 
will mainly see the merger and ring-down of the gravitational 
wave (GW) signal, which can now be computed numerically
using the full Einstein equations. Using numerical waveforms that
also include about ten GW cycles of inspiral, we improve inspiral-only
position estimates by an order of magnitude. We show that 
LISA localizes half of all 
such systems at $z=1$ to better than 3 arcminutes and the best 
20\% to within one arcminute. This  
will give excellent prospects for identifying the host galaxy.
\end{abstract}


\maketitle

\paragraph{Introduction.---}

Astronomical observation provides very strong evidence for the existence of
a ``dark'' compact massive ($10^6 - 10^9\, M_{\odot}$) object in the core of 
every galaxy for which the central parsec region can be resolved
\cite{Tremaine:2002js}.  These objects are believed to be supermassive black 
holes (SMBH), and are of great interest to researchers in
fundamental physics, astrophysics and cosmology.
Their formation and the observed correlation between SMBH mass and galaxy
morphology (see \cite{Merritt:2004gc} for an overview) are still open questions.
SMBHs probably arise at least in part through the mergers of smaller-mass BHs 
\cite{Sesana:2007sh}. These mergers and the mergers of SMBH binaries
following collisions of galaxies
constitute
some of the most powerful sources of gravitational waves
(GW) predicted by current models. We will be able to detect such events 
throughout the Universe with LISA \cite{LISA1reduced}, a proposed space-borne
gravitational-wave observatory, scheduled for launch in 2018+ and designed
to be sensitive to GW signals in the range 
$10^{-4} - 0.1\; \rm{Hz}$.  

Realizing the full scientific benefit of the LISA mission will require accurate
estimates of the binary's parameters.
Precise measurements of the masses, spins and distance will allow us 
to probe models of SMBH formation. Accurate localization of the source
in the sky is crucial to relate gravitational-wave and 
electromagnetic observations of the coalescence event, and hopefully will 
allow identification of the host galaxy. Optical observations are
required to measure the redshift to the object, while gravitational-wave
observations yield precise calibration-free estimates of the distance.
Such LISA events with optical counterparts will determine 
the redshift-distance relationship, which in turn
will allow us to map the geometry of the Universe and measure the
amount of dark energy. 

Recently several groups \cite{Lang:2007ge, Arun:2007hu, Trias:2007fp,
Porter:2008kn} have evaluated the accuracy of parameter estimation 
using the inspiral part of the GW signal. It was shown that the spin
and higher orbital harmonics are necessary to de-correlate the parameters  
and therefore improve the parameter estimation. 

In this Letter we assess the angular resolution of LISA for SMBH binaries with 
(red-shifted) masses above $10^7\, M_{\odot}$. We expect several such merger
signals per year at relatively close distance \cite{Sesana:2007sh}.
For those
heavy systems the inspiral signal may be smaller or at least not much larger
than the instrumental noise, and the signal will be dominated by the merger and ring-down. We use
numerical relativity to compute a waveform which contains about  
ten GW cycles, plus merger and ring-down. We fix the {\em redshifted} masses of two non-spinning BHs to 
$4.44\times 10^6\, M_{\odot}$ and $8.88 \times 10^6\, M_{\odot}$ and vary the ``extrinsic''
parameters: sky ecliptic coordinates $\theta_S,\, \phi_S$, inclination $\iota$ of the orbital 
angular momentum to the line of sight, 
polarization angle $\psi$, fiducial arrival time $T_0$ which we fix to
be the time when the binary separation equals $10\, M$, orbital phase
$\phi_0$ at $T_0$, and luminosity distance $D_L$. 

Data analysis for LISA is based on time-delay interferometry (TDI, see
\cite{Tinto:2004wu} for an overview). We convert the strain polarizations $h_{+}$
and $h_{\times}$ given in the source frame to first generation
unequal-arm Michelson streams $X,Y,Z$ \cite{Tinto:2004wu,Vallisneri:2004bn}
and use two combinations $A = (2X -Y -Z)/3$, $E =
(Z-Y)/\sqrt{3}$ with uncorrelated noise. 
Due to technical difficulties explained below, we do not
take into account the third independent combination,  
which has poor sensitivity to GW at low frequencies, and which adds only a few percent to
the combined signal-to-noise ratio (SNR). By
fixing masses we underestimate the errors, but at the same time not
including the third TDI combination overestimates the error boxes. We
mainly concentrate here on the estimation of the localization of the  
source and usually \cite{Porter:2008kn} the directional angles very weakly correlate with masses.
Since we use the merger for localizing the hosting galaxy, we cannot produce an early 
warning for the merger itself; however, we can identify the hosting galaxy by the afterglow 
\cite{Milosavljevic05}.

We conduct Monte Carlo simulations by randomly choosing 600 points in the extrinsic parameter space
(with fixed masses, and choosing an example distance of $z=1$ or $D_L \approx 6.4$ Gpc),
and estimate the errors in the parameter error box with two different methods: 
the first is based on computing the variance-covariance
matrix, the second is a Bayesian method based on the evaluation of
the marginalized posterior distribution function using a Metropolis-Hastings Markov chain (MHMC) 
\cite{Chib1995, Christensen:2001cr, Cornish:2006ms}.
We have shown that both methods give comparable results. For $50\%$ of the 
randomly chosen parameters we can localize the source down to a box 3 
arcminutes on a side, and the best 20\% of the events are localized to 
better than one arcminute. The relative error in the distance is less than
1\%,  the  largest error in the distance being due to weak
 lensing \cite{Holz:2005df}. 
We also estimate robustness of our results with respect to 
errors in the numerical waveform. By conducting another Monte
Carlo simulation, we have found that the errors in the sky locations  
 are good to within factor two at worst (but usually much better than that).

\paragraph{Numerical relativity waveforms.---}

We use numerical-relativity waveforms from simulations of non-spinning
black holes at mass ratio 1:2, which have initially been presented in
\cite{Ajith:2007kx}, and are discussed in more detail in
\cite{Damour:2008te}, where the quadrupole spherical harmonic mode has
been compared with effective-one-body waveforms, and excellent
agreement in the phase evolution has been found. Here we now include
results for higher spherical-harmonic contributions.
Our simulations follow the now standard moving-puncture approach
\cite{Campanelli:2005dd,Baker05a}, using the BAM 
code~\cite{Bruegmann:2006at,Husa2007a} to numerically solve the Einstein equations.
Black hole initial data are modeled as conformally flat puncture data \cite{Brandt97b,Hannam:2006vv}.
The initial data parameters then reduce to the black-hole masses, momenta and separation. We choose
a separation of 10 $M$ (we refer to the total initial black-hole mass as $M$). The momenta are
specified to give quasicircular inspiral with minimal eccentricity of $e \approx 0.003$ \cite{Husa:2007rh}. 

The gravitational-wave signal is extracted at five surfaces of constant radial
coordinate, $R_{ex} = 40, 50, 60, 80,  90 M$ by means of the Newman-Penrose Weyl
tensor component $\Psi_4$, as described in \cite{Bruegmann:2006at}.  
At every extraction radius the gravitational wave strain is obtained
from $\Psi_4$ by double time integration as described in \cite{Ajith:2007kx}.
The analysis carried out in this paper will use, as approximate asymptotic
amplitude, the curvature perturbation extracted at radius $90M$, at our highest
numerical resolution, as has been done in \cite{Damour:2008te}. 
For the modes with $l>2$ we filter out low frequencies and
frequencies higher than the Nyquist frequency to avoid the pathologies
discussed in Sec. II.A. of \cite{Berti:2007fi}.   
For these simulations, we find that the finite extraction radii
dominate the error, and the amplitude error is below 5\% prior  
to merger, and the accumulated phase error is below 0.25 radians for the $700M$ 
up to $M\omega = 0.1$. The fall-off in the amplitude error with
respect to radiation extraction radius is not clean around merger
time, preventing us from performing an accurate extrapolation to  
infinity. As such, we would conservatively give an uncertainty estimate of 10\%
of the amplitude at merger and later. We estimate the {\it relative}
error in the amplitude between different modes (which is what
dominates the results presented here) as below 4\%.

\paragraph{Parameter estimation.---}
\label{Estimations}

We have used two methods to estimate accuracy in measuring the parameters of the 
GW signal. The first method is based on computing the variance-covariance matrix.
This method is widely used and well described in numerous publications (see for example
 \cite{Cutler:1994ys, Cutler:1997ta, Vallisneri:2007ev,
   Holz:2005df}). It is based on inverting the Fisher matrix $\Gamma$, which is
 the matrix of the inner products between derivatives of the signal 
 with respect to the parameters $h_{,i} = \partial h/\partial \lambda_i$:
 
 \begin{equation}
 \Gamma_{ij} = 4\Re \int_0^{\infty} df\, \frac{\tilde{h}_{,i}(f)\tilde{h}^*_{,j}(f) }
 {S_n(f)},
 \end{equation}
where $S_n(f)$ is the one-sided noise power spectral density. 
We use two TDI measurements and the combined Fisher matrix is a sum of Fisher
matrices for $A$ and $E$: $\Gamma_{ij} = \Gamma^A_{ij} + \Gamma^E_{ij}$. 
The presence of the noise might cause a deviation of the recovered parameters of the GW 
signal from the true values. The diagonal elements of the
variance-covariance matrix are maximum likelihood estimators of the
variance of parameters around the true value in the case of a large
SNR (which is always the case here). We have computed derivatives
numerically using forward differencing and checked the robustness with
respect to the step using several randomly chosen points in the  
parameter space. We have tried to choose a parametrization of the signal which would reduce
the dynamical range of the eigenvalues of the Fisher matrix, however there are still 
six or seven orders of magnitude between the smallest and largest eigenvalues. 

As a second way to evaluate the parameter estimation error we 
use a Metropolis-Hastings Markov chain (MHMC) \cite{Chib1995} to
estimate the posterior probability function 
\begin{equation}
p(\vec{\lambda}|s) \propto \pi(\vec{\lambda}) L(s|\vec{\lambda}),
\end{equation}
 where $\pi(\vec{\lambda})$ is the prior distribution of parameters and $L(s|\vec{\lambda})$
 is the likelihood function. The MHMC uses a proposal distribution $q(\vec{\lambda}_{(i)}
 | \vec{\lambda}_{(j)})$ and Metropolis-Hastings ratio
 \begin{equation}
 H = \frac{p(\vec{\lambda}_{(j)}|s) q(\vec{\lambda}_{(j)}| \vec{\lambda}_{(i)})}
 {p(\vec{\lambda}_{(i)}|s) q(\vec{\lambda}_{(i)}| \vec{\lambda}_{(j)})},
 \end{equation}
 which gives a probability $\alpha(\vec{\lambda}_{(i)}| \vec{\lambda}_{(j)}) = min(1, H)$
 of accepting the jump from the point $\vec{\lambda}_{(i)}$ to $\vec{\lambda}_{(j)}$.
 MHMC gives the best result (better convergence) if the proposal
 distribution matches the shape of the target distribution. For the
 proposal distribution we take normal jumps in the  
 eigen-directions of the Fisher matrix \cite{Cornish:2006ms}, yielding 
 acceptance rates $> 30\%$. We have generated simulated data using
the {\it{lisatools}} software [http://code.google.com/p/lisatools/],
 consisting of instrumental noise and a simulated reduced Galaxy confusion noise with
 more than $5\times 10^7$ chirping binaries \cite{MLDC3,Cornish:2007if}. 
 We have generated signals for 600 randomly chosen parameters (600
 data sets), and performed 600  mappings of posterior distribution
 functions using $3\times 10^5$ long chains. We have also 
 used mild simulated annealing for the first two thousands steps,
 which helps the chain to migrate from the point with true parameters
 to the point with better likelihood. Note that due to the presence of
 the Galaxy the noise is strictly speaking not Gaussian. Because the
 jumps are rather small, we have made the assumption that the Fisher matrix
 does not change notably within the jumping region and therefore 
 $q(\vec{\lambda}_{(i)}| \vec{\lambda}_{(j)}) =
q(\vec{\lambda}_{(j)}| \vec{\lambda}_{(i)})$ and the
Metropolis-Hastings ratio is reduced to Metropolis. This assumption
results in a small bias in the estimation
of the variance, but it significantly speeds up the computations.  

\begin{figure}[ht]
\includegraphics[height = 0.29\textheight, keepaspectratio=true]{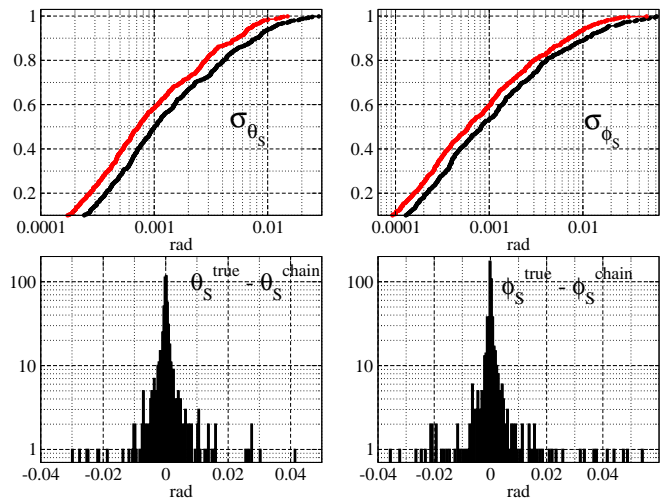}
\caption{Top plots: cumulative histograms for the standard deviation
  estimators from 600 randomly chosen points in the parameter
  space. 
  The black (lower) curve corresponds to the variance $\sigma$
  evaluated by computing the variance-covariance matrix and the red (upper) curve
  is the Bayesian estimator obtained using the MHMC. 
Lower plots: deviation of the mean value of the chain from
  the true parameter of the signal. 
The plotted results correspond to a distance of $6.4$ Gpc ($z \approx 1$ and
total mass of $\approx 0.7\times 10^7\, M_{\odot}$),
} 
\label{Hist}
\end{figure}

The results of both methods are presented in the top two plots of
Figure~\ref{Hist}, as cumulative histograms for the 600
realizations.  
Both methods give comparable 
estimation of the error box for the sky location $(\theta_S, \phi_S)$. One can see that 50\% of  the trials give an error box 
smaller than 3 arcminutes for a source located at $z=1$.
 
The duration of the signal from a separation of $10 M$ to the merger is 26 hours,
so the Doppler modulation is small, however for our choice
of parameters in the Monte Carlo simulation the SNR varies between 900 and 9000, and
these large values
drastically improve parameter estimation. The other crucial ingredient
for the excellent angular resolution is the presence of higher
orbital harmonics coming from the higher multipoles of the source. We
find that the use of only the dominant ($l=\vert m\vert =2$) mode worsens
our results by up to a factor of ten. Different harmonics have
different angular emission patterns and different dependencies on the
inclination. These help to de-correlate the parameters, and together
with the high SNR turn a seemingly small effect (higher modes are much lower in
amplitude) into a crucial contribution for parameter estimation.
For our analysis we have used $l=2,3,4$ and $m=-l,\ldots l$ (except $m=0$).
We also find that the arrival time could be measured with an
accuracy of less than a second and the luminosity distance with an
accuracy less than 1.5\%. However, 
the dominant error in
estimating the luminosity distance is due to weak lensing \cite{Holz:2005df}.
Weak lensing gets stronger with distance \cite{Holz:2004xx} and we expect to have 
SMBH binaries at distances $z<5$.   

As we have mentioned, the Markov chain is usually migrating to the point with higher 
likelihood and this shift could be as large as the variance itself.  We have found
 that the secondary maxima in the likelihood could be very close  to the primary maximum  
 both in amplitude and 
 distance in parameter space. In our simulations we observed 
 that due to the presence of the instrumental and Galactic confusion noise the secondary 
 maxima could become stronger than the primary (by less than 0.01\%) and the secondary 
 could be located about a $\sigma$ away, where $\sigma$ is the
 standard deviation of the parameter assuming no secondary maxima. In
 the lower two plots of Figure~\ref{Hist} we show the histogram of the
 deviation mean value of the chain from the true sky location. Further work 
is needed to explore ways of reducing these ambiguity errors.

We have used only two TDI streams in the above analysis, because we could
not compute the third independent data set constructed out of
$X,\,Y,\,Z$, which is $T \propto X + Y + Z$, with the required accuracy:
the signal cancellation at low frequencies was not perfect due to
inaccuracy (less than a few percent) in the evaluation of $X,\,Y,\,Z$. In
principle we could generate $A,\, E,\, T$ out of the six-pulse TDI
combinations as in \cite{Prince:2002hp} but we did not 
have simulated a Galactic background for those combinations, so we have decided to drop the 
third effective detector for the present considerations. We however checked that using
$T$ (as in \cite{Prince:2002hp}) would improve our SNR by a few percent, and, more
importantly, could also improve parameter estimation: having three independent 
measurements should help to triangulate the source.

Finally, we have checked the robustness of the variance estimation with respect to 
possible errors in the numerical waveform. Since the results depend on
the relative strengths of the different harmonics, we varied the
$l=m=2$ mode by $\pm$5\% with respect to the higher
harmonics.  We have estimated the variance by computing the
variance-covariance matrix and comparing it to the original one, and find that
$\sigma$ changes by 
at most a factor of two (much less in the majority of the cases $\lesssim
15\%$). We have also noticed that changes in the waveform resulting
from enhancing the higher orbital modes improve 
results (makes the variance smaller) which is in agreement with our explanation of the angular 
resolution.

\paragraph{Summary.---}
\label{sum}

We have presented a first study of how results from numerical relativity can
improve parameter estimation for SMBH binary observations with LISA. We have
chosen to first examine the issue of angular resolution of LISA -- which is 
crucial to identify electromagnetic counterparts to gravitational wave observations. 
Looking at the merger signal for non-spinning SMBH binaries with redshifted total mass 
$1.4\times 10^7\, M_{\odot}$
and a mass-ratio of 2,
Placing our source at a luminosity distance of $6.4$ Gpc ($z \approx 1$, corresponding 
to a total mass of $\approx 0.7\times 10^7\, M_{\odot}$), 
we have found an excellent sky resolution: for $50\%$ of our
randomly chosen events we can localize the source down to 3 arcminutes, which roughly  corresponds to an order of magnitude improvement over
estimates using only the inspiral phase of the GW signal.
We have also shown that we obtain consistent error estimates when
comparing two independent methods based on the variance-covariance matrix
and a Metropolis-Hastings Markov chain.

Our result calls for further work along several directions. First, it is
necessary to cover the parameter space of numerical simulations: larger mass ratios
will show two effects: a decrease in SNR, but an increase of the contributions of higher modes,
which have proven crucial for parameter estimation.
The inclusion of spins will lead to a much larger parameter space, but also to much more
interesting waveform structures. For lower masses, it will be important to accurately match
numerical waveforms to post-Newtonian results in order
to cover the whole LISA band, and to understand the systematical errors in the matching procedure.
The extension of existing phenomenological waveforms (e.g. along the lines of \cite{Ajith:2007qp,Ajith:2007kx})
to spinning binaries and non-dominant modes will allow for systematic parameter estimation
studies in the regime where solving the full Einstein equations is necessary to obtain accurate
gravitational wave signals, and will extend the useful sensitivity range of LISA to substantially higher masses,
so that LISA can begin to explore the process whereby black holes grow from the size we see in our galaxy to the sizes
that are needed in quasars.

\paragraph{Acknowledgments. ---}
We have benefitted from discussions with Alicia Sintes.
S.~H.~is a VESF fellow of the European Gravitational Observatory
(EGO).
M.~H.~was supported by FSI grant 07/RFP/PHYF148, and thanks
AEI Potsdam for hospitality.
This work was supported in part by DFG grant SFB/TR~7
``Gravitational Wave Astronomy'' and DLR (Deutsches Zentrum f\"ur
Luft- und Raumfahrt).
Computations were performed at the Morgane cluster at AEI Potsdam,
original numerical relativity simulations were performed at
LRZ Munich and  at Theoretisch-Physikalisches Institut, 
FSU Jena.


\bibliography{refs}


\end{document}